# Comprehensive Framework for Evaluating Conversational AI Chatbots

Shailja Gupta*, Carnegie Mellon University, USA

Rajesh Ranjan*, Carnegie Mellon University, USA

Surya Narayan Singh*, BIT Sindri, India

**Abstract**: *Conversational AI chatbots are transforming industries by streamlining customer service, automating transactions, and enhancing user engagement. However, evaluating these systems remains a challenge, particularly in financial services, where compliance, user trust, and operational efficiency are critical. This paper introduces a novel evaluation framework that systematically assesses chatbots across four dimensions: cognitive and conversational intelligence, user experience, operational efficiency, and ethical and regulatory compliance. By integrating advanced AI methodologies with financial regulations, the framework bridges theoretical foundations and real-world deployment challenges. Additionally, we outline future research directions, emphasizing improvements in conversational coherence, real-time adaptability, and fairness.*

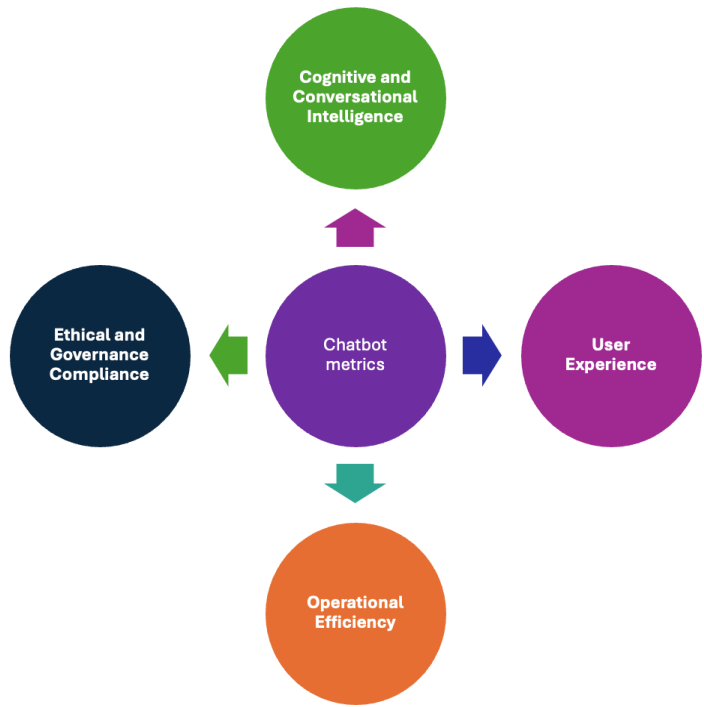

Figure 1: Four themes of metrics for holistic evaluation of Chatbots

I. Introduction

The services industry has increasingly relied on **conversational AI**, particularly chatbots, to automate customer interactions, deliver personalized advice, and perform secure transactions. N**atural language processing (NLP)** and **machine learning (ML)** have enabled these systems to handle complex tasks, from answering customer inquiries to executing transactions. However, evaluating the effectiveness of chatbots in financial services requires a more specialized framework that addresses the industry's unique needs, including accuracy, compliance with regulations, and customer trust.

This paper proposes a comprehensive framework for evaluating financial chatbots, covering four key areas:

- **Cognitive and conversational intelligence**: the chatbot's ability to understand, retain context, and engage in multi-turn conversations.

- **User Experience**: How well the chatbot satisfies users' needs and engages them in meaningful conversations.

- **Operational Efficiency**: The system can handle a high volume of interactions while maintaining speed and cost-effectiveness.

- **Ethical and Governance Compliance**: Ensuring the chatbot complies with financial regulations and adheres to fairness and transparency standards.

The framework is grounded in existing AI theories, including **reinforcement learning (RL)** (Sutton et. al., 1998), **contextual embeddings** (Vaswani et. al., 2017), and **fairness-aware AI** (Dwork et. al., 2011). Furthermore, this paper highlights the **novelty** of integrating domain-specific metrics with the theoretical underpinnings of AI, a gap often overlooked in more generic chatbot evaluations. Here domain is considered considered financial industry as this industry has a high security and trust requirement. However, these metrics apply to other industries as well.

II. Related Work

Explainable AI (XAI) in Chatbots: Explainability is crucial in financial services, where AI systems must be transparent and interpretable to regulators, customers, and business stakeholders. **Explainable AI (XAI)** techniques such as **SHAP** (Lundberg et. al., 2017) and **LIME** (Ribeiro et. al., 2016) allow AI systems to provide insights into decision-making processes. While these techniques are computationally expensive, their importance in financial chatbots cannot be overstated, especially when regulatory compliance is a factor.

AI Governance and Ethical Compliance: In financial services, chatbots must comply with regulations such as the **General Data Protection Regulation (GDPR)** (Voigt et. al., 2017), **Payment Services Directive 2 (PSD2)**, and anti-money laundering laws. **AI governance** frameworks ensure that these systems adhere to legal standards for data privacy, fairness, and transparency (Arrieta et. al., 2019). However, there is limited research that combines these governance frameworks with advanced AI models in production environments, which this paper addresses by recommending a set of metrics to be measured for production ChatBots.

Cognitive Models and Conversational AI: Recent advancements in **transformer models** such as **BERT** (Devlin et. al., 2018) and **GPT** have significantly improved chatbots' natural language understanding (NLU) capabilities. These models leverage self-attention mechanisms [9] to retain context over long conversations, which is critical in financial interactions where users may engage in multi-turn dialogues (e.g., loan applications and investment queries). This has enriched the conversation that happens between bots and users, which further necessitates the need for a **comprehensive evaluation framework** that integrates these cutting-edge AI models with metrics tailored for the financial industry.

III. Proposed Framework

The proposed framework evaluates conversational chatbots across four categories. Each category is grounded in theoretical AI principles and adapted to meet the specific needs of financial services.

A. Cognitive and Conversational Intelligence:

**Cognitive and conversational intelligence** refers to the chatbot's ability to understand user inputs, maintain context across multiple interactions, and provide accurate and relevant responses. Financial chatbots are often tasked with handling complex transactions, such as loan applications, financial advice, and fund transfers, where accurate understanding and retention of context are critical. The following metrics are essential for evaluating the chatbot's cognitive performance. The theoretical grounding for this is provided by **statistical models of language** such as **transformers** (Sutskever et. al., 2014) and **sequence-to-sequence models** (Boyd et. al., 2004).

Key Metrics that Capture the Essence of Cognitive and Conversational Intelligence:

**Natural Language Understanding (NLU) Accuracy** measures the chatbot's ability to correctly interpret and classify user intents. In financial services, this is crucial for tasks like identifying whether a user is requesting a balance check, making a fund transfer, or asking for investment advice.

$$NLU\ Accuracy\ =\ \frac{1}{n}\sum_{i=1}^{n} I(\widehat{y_i} = y_i)$$

$\widehat{y_i}$: predicted intent by the chatbot.
$y_i$: actual intent provided in the labeled data.
$n$: total number of user queries.

NLU accuracy directly correlates with the chatbot's ability to **understand user intents** correctly. This metric is particularly relevant in financial services because a misunderstanding of user intent can lead to serious consequences, such as incorrect financial advice or failed transactions. High NLU accuracy ensures that the chatbot is reliably identifying user requests, which leads to more effective task completion and customer satisfaction. Additionally, **the F1-Score** can be used to provide a balanced measure of precision and recall in intent classification.

$$F1\ Score\ =\ \frac{2*Recall*Precision}{Recall+Precision}$$

The **F1-Score** helps assess the chatbot's ability to correctly identify intents in a balanced way, especially in cases where the data may be imbalanced (e.g., when certain intents occur more frequently than others).

**Context Retention** measures how well the chatbot maintains the conversational context across multiple turns. In financial chatbots, context retention is crucial for handling multi-step processes, such as completing a loan application or executing a series of financial transactions.

$$Context\ Retention\ =\ D_{KL}\ (P(C_{t-1})||P(C_t))$$

$D_{KL}$: The **Kullback-Leibler Divergence**, a measure of how one probability distribution (the context at the previous turn $C_{t-1}$) diverges from the updated context distribution $C_t$. For financial chatbots, maintaining context over multiple turns is critical for complex interactions. For instance, if a user asks for their account balance and then immediately asks to transfer funds, the chatbot must remember the previous context to complete the task without repeating unnecessary steps. Low divergence (low $D_{KL}$) indicates that the chatbot retains context effectively across turns, ensuring that users don't have to re-explain their queries. Essentially, low divergence implies that the chatbot does not "forget" or deviate from the previous interaction, leading to more fluid and human-like conversations. This leads to smoother, more human-like conversations, improving the overall user experience.

**Conversational coherence** evaluates whether the chatbot's responses logically follow the user's queries. In financial services, coherence is essential, as users may ask a series of related questions (e.g., account status, transaction history, and fund transfers), and the chatbot must provide logically connected responses. The cost of giving wrong answers can not only impact the brand image but can also lead to legal hassles. A coherent conversation ensures that the chatbot's responses follow a logical flow, which is especially important in **multi-turn dialogues** that involve multiple, interdependent financial operations. For example, if a user asks about their loan status and follows up with a question about the next payment date, the chatbot's response should maintain logical coherence and accuracy throughout. This can be evaluated qualitatively or with metrics like **BLEU (Bilingual Evaluation Understudy Score)**:

$$BLEU\ Score\ =\ BP\ *\ exp\ (\sum_{n=1}^{N} w_n\ log\ p_n)$$

$p_n$: The precision of n-grams.
$w_n$: The weight assigned to each n-gram.
$BP$: Brevity penalty (to discourage overly short responses).

A higher **BLEU Score** indicates that the chatbot is providing coherent, fluent responses that are close to human-level responses, making it an important metric for assessing conversational quality in complex financial interactions.

**Task Completion Rate (TCR)** measures the percentage of tasks successfully completed by the chatbot. In financial services, this includes tasks such as transferring funds, applying for loans, updating account details, and providing accurate financial advice.

$$TCR = \frac{Completed\ Tasks}{Total\ Initiated\ Tasks}$$

This metric is one of the most direct indicators of chatbot effectiveness in performing real-world tasks. For financial chatbots, **high TCR** is essential, as users expect chatbots to perform tasks such as fund transfers, bill payments, or investment planning without errors or delays. A low TCR suggests that the chatbot is failing to complete tasks correctly, which can lead to user frustration and distrust in the system. Task completion also reflects the chatbot's capability to handle complex financial interactions end-to-end. For instance, a user applying for a loan might interact with the chatbot multiple times during the process, so a high TCR would indicate that the chatbot is capable of guiding the user through the entire process seamlessly.

**Semantic Similarity Score** (Gupta et. al., 2024) measures how semantically close the chatbot's response is to the ideal response. This is critical in financial services, where the chatbot must not only provide grammatically correct responses but also deliver **accurate and relevant** information.

$$Semantic\ Similarity = cosine\ similarity(R_{Bot},\ R_{Ideal})$$

$R_{Bot}$: Vector representation of the chatbot's response.

$R_{Ideal}$: Vector representation of the ideal response.

The **Semantic Similarity Score** assesses the degree to which the chatbot's response aligns with the expected (or ideal) response in terms of meaning. For example, if a user asks for financial advice, the chatbot should not only provide a grammatically correct response but also one that is factually accurate and relevant. A high Semantic Similarity Score indicates that the chatbot is providing semantically accurate information, which is vital in regulated financial contexts where incorrect advice could have serious consequences.

**Turn-Taking Balance** measures the proportion of user turns versus chatbot turns in a conversation, which can reflect how interactive and engaging the chatbot is.

$$Turn - Taking\ Balance = \frac{ChatBot\ Turns}{Total\ Turns}$$

In natural, human-like conversations, there should be a balance between the chatbot and the user contributing to the conversation. A **balanced turn ratio** suggests that the chatbot is both responsive to the user's inputs and proactive in guiding the conversation. For financial chatbots, a balance helps ensure that the chatbot is not dominating the conversation or requiring too much input from the user, which could frustrate them during complex tasks such as financial planning or applying for a loan.

B. User Experience

In the financial services sector, **user experience (UX)** is crucial for maintaining customer satisfaction and driving long-term retention, especially in a competitive landscape where customers expect seamless, accurate, and personalized interactions. Positive user experiences lead to stronger customer loyalty, increased recommendations, and higher lifetime value. Drawing from **utility theory** (Von et. al., 1944) and **engagement entropy** (Shannon et. al., 1948), this category focuses on evaluating how well the chatbot satisfies users' needs and keeps them engaged in meaningful conversations.

Key Metrics that Capture the Essence of User Experience:

**Customer Satisfaction (CSAT)** captures user satisfaction after interacting with the chatbot. It's typically collected through post-interaction surveys where users rate their experience on a scale (e.g., 1 to 5 or 1 to 10).

$$CSAT = \frac{1}{n}\sum_{i=1}^{n} S_i$$

$S_i$: The satisfaction score given by user i after interacting with the chatbot.
$n$ : The total number of respondents.

CSAT is a **direct measure of user sentiment** after interacting with the chatbot. It reflects how satisfied users are with the chatbot's ability to meet their needs, whether it's providing quick answers, solving financial problems, or offering helpful advice. In financial services, a high CSAT score suggests that the chatbot is effective in delivering a positive experience, which is critical for customer retention. To improve CSAT, chatbots need to excel at delivering **timely, accurate, and contextually relevant responses**. When users feel that the chatbot has understood their queries and provided helpful assistance, they are more likely to give a high satisfaction score.

**Net Promoter Score (NPS)** measures the likelihood of users recommending the chatbot service to others. It's calculated by subtracting the percentage of detractors (those who are unhappy with the service) from the percentage of promoters (those who are highly satisfied).

$$NPS = \frac{(\eta_{Promoters} - \eta_{Detractors})}{\eta_{Total}} * 100$$

**Promoters** are users who rate their likelihood to recommend the service as high (typically 9-10 on a 0-10 scale). **Detractors** are users who rate their likelihood to recommend the service as low (typically 0-6 on a 0-10 scale). NPS is a **long-term indicator** of user loyalty and satisfaction. A high NPS score suggests that users not only find the chatbot helpful but are also likely to recommend it to others. In the financial services sector, where trust and reliability are critical, a strong NPS score can translate into **brand loyalty** and **customer advocacy**. By tracking NPS, financial institutions can gauge whether the chatbot contributes positively to the overall user experience. A chatbot that provides helpful financial advice, solves issues quickly, and personalizes interactions will encourage users to become promoters.

**Engagement Depth** measures the complexity and depth of user engagement with the chatbot by counting the number of conversational turns in an interaction. This metric reflects how deeply users interact with the chatbot, rather than just initiating superficial queries.

$$ED = \frac{1}{n}\sum_{i=1}^{n} T_i$$

$T_i$: The number of conversational turns in the iii-th interaction.

$n$: The total number of interactions.

**Engagement Depth** is particularly important for financial chatbots, as it indicates how effectively the chatbot engages users in **complex, multi-turn conversations**. For example, when helping users with financial planning, loan applications, or investment queries, the chatbot may need to ask follow-up questions and guide the user through a series of steps. A higher **engagement Depth** suggests that users are interacting more fully with the chatbot, engaging in detailed conversations that likely indicate the chatbot's ability to handle more complex queries. However, excessive engagement depth could also be a sign of inefficiency (if the chatbot requires many turns to resolve simple issues), so this metric must be interpreted in context with other performance indicators like **Task Completion Rate (TCR)**.

**Cumulative Utility Gain**: measures the total utility (or value) that a user derives from interacting with the chatbot over a series of interactions. This metric is derived from **utility theory** [11], which posits that users seek to maximize the benefit they receive from a service or interaction.

$$U(x) = E\left[\sum_{i=1}^{n} u(T_i)\right]$$

$u(T_i)$: The utility derived from the i-th interaction, where $T_i$ represents the number of tasks completed.

$n$: The total number of interactions.

**Cumulative utility gain** is a theoretical measure of the **overall benefit** a user receives from interacting with the chatbot over time. In financial services, users may return to the chatbot for multiple interactions (e.g., checking account balances, transferring funds, applying for loans). The goal is to maximize the user's perceived value from these interactions by ensuring that the chatbot is **efficient, accurate, and helpful** across all touchpoints. This metric captures the **long-term value** the chatbot provides, reflecting how well it helps users achieve their financial goals, solves problems, and improves their overall experience. A high **Cumulative Utility Gain** means that the chatbot consistently delivers meaningful and valuable interactions, which enhances customer satisfaction and retention.

C. Operational Efficiency

In financial services, the efficiency with which a chatbot operates is paramount. Financial institutions often deal with thousands, if not millions, of customer interactions daily. Chatbots must respond quickly and accurately while being able to scale according to demand. **Queueing theory** (Kleinrock et. al., 1975) and **resource optimization** (Boyd et. al., 2004) provide the theoretical basis for understanding and measuring the efficiency of these systems. Queueing theory models how requests (queries) are processed in a system with limited resources, while resource optimization ensures that these resources are allocated in a way that minimizes costs while maintaining performance.

Key Metrics that capture the essence of operational efficiency:

**Average Response Time (ART)** measures the average time taken by the chatbot to respond to user queries. This is one of the most critical metrics for **user experience** and operational performance, particularly in financial services, where users expect near-instant responses to their inquiries.

$$Average\ Response\ Time\ (ART) = \frac{1}{n}\sum_{i=1}^{n} RT_i$$

$RT_i$: The response time for the iii-th query.

$n$: The total number of queries during the measurement period.

A **short response time** is crucial for maintaining user engagement and satisfaction. In financial contexts, users expect chatbots to handle complex queries, such as checking account balances, transferring funds, or providing investment advice, in real-time. A delay of even a few seconds can result in a poor user experience, especially if the user is in a time-sensitive situation (e.g., stock trading or paying bills). The **average response time** (ART) is influenced by several factors, including the complexity of the query, system load, and the efficiency of the chatbot's backend processing system. Optimizing ART involves balancing the **scalability** of the system (i.e., handling multiple queries at once) with the **processing power** available. Lowering the ART can significantly improve user satisfaction and engagement, making this a key operational metric. This metric can be optimized by improving **resource allocation** during peak loads using methods such as **dynamic load balancing** (Kleinrock et. al., 1975).

**Automation Rate** measures the proportion of queries handled autonomously by the chatbot without the need for human intervention. This is an important indicator of the chatbot's effectiveness in reducing the workload on human agents, particularly in large-scale financial operations.

$$Automation\ Rate = \frac{Automated\ Queries}{Total\ Queries}$$

Automated Queries refer to the number of queries resolved entirely by the chatbot without human involvement, and Total Queries refer to the total number of queries received. A high **automation rate** indicates that the chatbot is capable of handling a large percentage of user inquiries independently, which can significantly reduce operational costs and improve efficiency. In the financial services industry, many customer interactions—such as balance checks, password resets, or simple loan inquiries—can be automated without needing human intervention. However, there are limits to full automation, especially for complex financial tasks that require human expertise, such as customized investment advice or complicated loan approvals. Thus, while maximizing the automation rate is desirable, the system must still provide **seamless hand-offs** to human agents when necessary. Optimizing the **automation rate** improves the scalability of the system, as fewer human resources are required to manage increasing volumes of customer interactions. This also improves overall **cost-efficiency**, which is a critical factor for large financial institutions that deal with high volumes of queries daily.

**Cost per Interaction (CPI)** tracks the cost-efficiency of each chatbot interaction, calculating the operational cost incurred by the financial institution for each query handled by the chatbot.

$$CPI = \frac{Total\ Interactions}{Operational\ Costs}$$

**Total Operational Costs** refer to the total cost associated with running the chatbot, including infrastructure, maintenance, and personnel costs, and **Total Interactions** refer to the total number of interactions handled by the chatbot over the measurement period. In financial services, controlling costs is a key objective, especially for institutions that handle millions of customer interactions daily. The **cost per interaction (CPI)** provides a direct measure of how efficiently resources are being used. Lowering the CPI means that the institution can handle more queries for the same or lower cost, improving the return on investment (ROI) of the chatbot system. CPI includes all **operational expenses**, such as server costs, software licensing fees, development and maintenance costs, and personnel involved in monitoring or updating the chatbot. If a chatbot can handle more queries autonomously (as indicated by the **automation rate**) while keeping costs low, the CPI will decrease, indicating higher cost efficiency. This metric is vital for determining whether the chatbot is a cost-effective solution, especially when compared to traditional human-powered customer support. In financial services, where margins can be tight, optimizing **CPI** can lead to significant savings while maintaining or improving service quality.

**System Uptime** tracks the percentage of time the chatbot system is operational and available to handle queries. In financial services, **high availability** is crucial, as downtime can lead to missed transactions, customer dissatisfaction, and potential financial losses.

$$System\ Uptime = \frac{The\ actual\ number\ of\ hours\ the\ chatbot\ was\ available\ and\ operational}{/the\ total\ number\ of\ hours\ the\ chatbot\ was\ expected\ to\ be\ available.}$$

**High availability** (often measured as **uptime**) is essential for financial chatbots because customers may need assistance or wish to complete transactions at any time, especially given the 24/7 nature of online banking and financial services. Even a short period of downtime can result in lost business opportunities, missed payments, or unsatisfactory customer experiences. Financial institutions typically aim for an **uptime** of 99.9% or higher, meaning that the system is available for nearly all scheduled hours. Achieving this involves having robust **fault-tolerant systems**, including backup servers and real-time monitoring to quickly identify and address any issues. The **system uptime** metric ensures that the chatbot remains available to handle queries, particularly during peak times, such as stock market hours or major financial events. Downtime must be minimized to avoid disrupting services, as these disruptions can lead to regulatory penalties or loss of customer trust.

D. Ethical and Governance Compliance

In the financial industry, ensuring that AI systems, especially chatbots, adhere to **ethical principles** and comply with **regulatory standards** is of utmost importance. Financial chatbots often handle sensitive data, provide personalized financial advice, and make decisions that can have significant consequences for individuals. Hence, their decisions must be **fair**, **transparent**, and in line with **legal regulations**. The framework incorporates **demographic parity** (Dwork et. al., 2011) and **fairness-aware AI** (Ranjan et. al., 2024; Hardt et. al., 2016).

Key Metrics that capture the essence of **ethical and governance compliance:**

The **bias detection rate** measures the chatbot's ability to detect and mitigate bias in its decisions. Bias can emerge in various forms, such as gender, racial, or socioeconomic biases, and can lead to unfair treatment of certain demographic groups, particularly in financial decisions like loan approvals, credit scoring, or investment advice.

$$Bias\ Detection\ Rate\ =\frac{Bias\ Free\ Decisions}{Total\ Decisions}*100$$

**Bias-free decisions** refer to the number of decisions or interactions in which the chatbot has not displayed any bias against a particular demographic group. **Total Decisions** refers to the total number of decisions or interactions processed by the chatbot. This metric evaluates how well the chatbot identifies and eliminates biased patterns from its decision-making processes. For instance, if the chatbot is disproportionately rejecting loan applications from a particular demographic group, this bias should be detected and rectified. A high **bias detection rate** indicates that the chatbot can make fair decisions across different groups without favoring or discriminating against any particular category. Bias detection is essential for financial systems, where legal and ethical implications are significant. For example, **fairness-aware AI models** can help mitigate biases by ensuring that the decisions are consistent across diverse user profiles without systematic discrimination (Dwork et. al., 2011; Hardt et. al., 2016).

**Compliance Rate** tracks how well the chatbot adheres to financial regulations, such as the General Data Protection Regulation (GDPR), Payment Services Directive 2 (PSD2), or anti-money laundering (AML) laws.

$$Compliance\ Rate\ =\frac{Compliant\ Interaction}{Total\ Interactions}*100$$

**Compliant interactions** refer to several interactions where the chatbot complies with relevant financial and data protection regulations, and **Total interactions** refer to the total number of interactions processed by the chatbot. In financial services, compliance with legal frameworks like **GDPR** (which governs data privacy in the European Union) or **AML** (which requires financial institutions to prevent money laundering) is critical. This metric assesses the chatbot's ability to meet these regulatory requirements during interactions. A high **compliance rate** shows that the chatbot is handling sensitive user data appropriately, following regulations, and not violating privacy standards. For example, the chatbot must not store user data without proper consent or share it without adhering to regulations. Financial organizations are liable for breaches of compliance, and this metric ensures that the chatbot aligns with regulatory standards (Voigt et. al., 2017).

**The explainability Score** measures the chatbot's ability to **transparently explain** its decision-making processes. Explainability is particularly crucial in financial services, where users and regulators need to understand how decisions are made, especially for high-stakes financial matters like credit approvals, loan recommendations, and risk assessments.

$$\text{Explainability Score} = \frac{\text{Explainable Decisions}}{\text{Total Decisions}}$$

**Explainable decisions** refer to the number of decisions where the chatbot can provide a clear, understandable explanation for its action or recommendation, and **total decisions** refer to the total number of decisions made by the chatbot. Explainability is a fundamental principle of **Explainable AI (XAI)**, ensuring that AI systems are transparent and understandable to users. In the financial industry, regulators and users often require an explanation for how a chatbot arrived at a particular decision. For instance, if a user is denied a loan, the chatbot must explain the reasoning behind the decision (e.g., based on the user's credit score, income level, or financial history). This metric ensures that the chatbot is not a "black box" but rather a system that can explain its decision-making process in a way that is comprehensible to non-experts. **SHAP** (Shapley additive explanations) and **LIME** (Local interpretable model-agnostic explanations) are common methods used to generate explanations in AI models. A high **Explainability Score** indicates that the chatbot is offering sufficient transparency in its decision-making.

**Fairness Parity Score** measures how equitable the chatbot's decisions are across different demographic groups. This metric uses **Wasserstein distance** to compare decision distributions between two or more groups, ensuring that there is **demographic parity**.

$$W(P1, P2) = \inf_{\gamma \in (P1, p2)} E[\|x1 - x2\|]$$

$W(P1, P2)$: The **Wasserstein distance** between two distributions $P1, and\ P2$ represents the outcomes for two demographic groups (e.g., Group 1: men, Group 2: women).
$E[\|x1 - x2\|]$: The expected distance between the two outcome distributions.

The **Fairness Parity Score** ensures that the chatbot's decisions are fair and unbiased across demographic groups. In financial contexts, unfair biases may arise when certain groups (e.g., minorities, women, or low-income individuals) consistently receive worse outcomes compared to others. For example, if a chatbot's lending decisions are biased against a particular demographic, this metric will help identify and quantify the disparity. The **Wasserstein distance** measures how different two outcome distributions are, and a lower value suggests closer parity between groups. A zero value means that the chatbot's decisions are completely fair and unbiased, as the decision outcomes for both groups are identical. This metric is vital in promoting fairness in AI, ensuring that financial chatbots do not propagate harmful biases (Gupta et. al., 2024).

The **novelty** of this framework lies in its **holistic approach** to chatbot evaluation in financial services, integrating advanced **AI models** with **industry-specific requirements**. The framework incorporates **regulatory compliance** metrics that are tailored for the financial industry, ensuring chatbots meet legal standards. Metrics like **market sensitivity** and **regulatory adaptation rate** highlight the need for chatbots to adjust dynamically to market changes and regulatory updates, areas largely ignored in generic chatbot evaluations. The framework bridges **theoretical models** of AI (such as reinforcement learning and fairness-aware AI) with practical metrics, creating an actionable set of tools for chatbot developers and financial institutions.

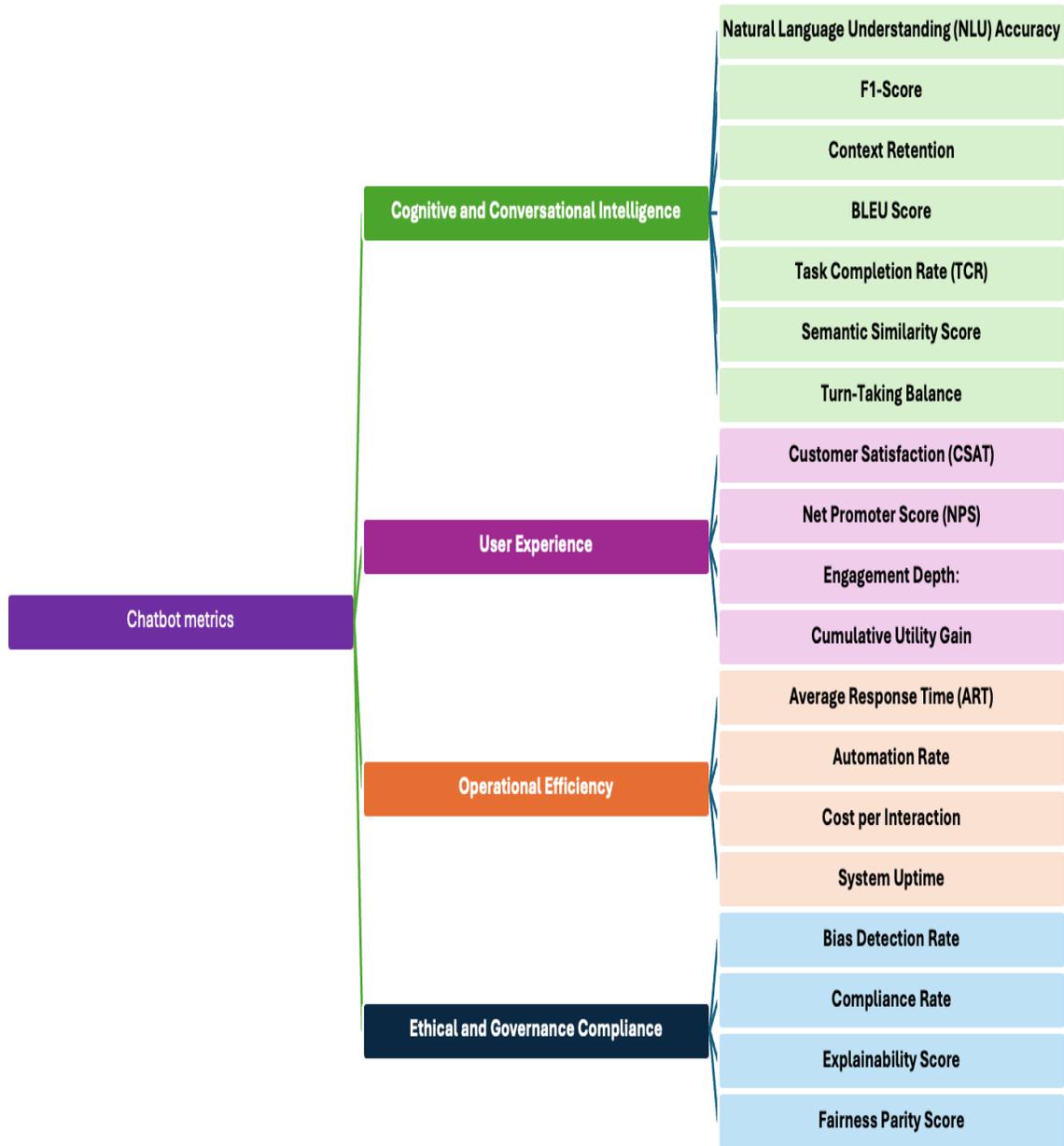

Figure 2: Summary of holistic evaluation metrics of ChatBot.

**V. Future Research Directions**

While the current framework provides a robust foundation, several areas for future research emerge:

**Enhancing Conversational Coherence:** Future work should focus on developing **hybrid models** that combine **transformers** and **reinforcement learning** to improve long-term conversational coherence.

These models will better handle the complexity of multi-turn financial conversations, ensuring that chatbots maintain accurate context and provide relevant responses across long dialogues.

**Real-Time Adaptability:** Improving the chatbot's ability to **adapt in real-time** to regulatory changes and market conditions is critical. Research into **meta-reinforcement learning** could enable chatbots to dynamically adjust their strategies in response to evolving financial landscapes.

**Fairness and Bias Detection:** Future research should also explore more **granular fairness metrics**, such as **intersectional fairness**, which considers multiple demographic dimensions (e.g., age, gender, income level) (Gupta et. al., 2024). This would ensure that chatbots provide equitable outcomes across all user groups in financial services.

**Privacy-Preserving AI:** As data privacy becomes more crucial, research into **privacy-preserving techniques** like **federated learning** and **differential privacy** is essential. These methods allow chatbots to improve performance without compromising user data privacy.

## VI. Conclusion

This paper presents a **comprehensive and novel framework** for evaluating conversational chatbots in the financial industry. By integrating advanced AI models with industry-specific requirements, the framework provides actionable metrics for assessing chatbot performance across cognitive, operational, user experience, and compliance dimensions. The proposed metrics and future research directions aim to guide the continuous improvement of chatbots, ensuring they remain efficient, secure, and compliant in the dynamic financial landscape.

*core contributors

Note: The university mentioned is the university from which the author has graduated